\documentclass[pra,onecolumn, showpacs, showkeys, secnumarabic, aps, amsmath, amssymb, nofootinbib, superscriptaddress, longbibliography, floatfix, table-of-contents, dblfloatfix]{revtex4-2}

\usepackage[utf8]{inputenc}
\usepackage[pdftex]{graphicx}
\usepackage{graphicx}
\usepackage{mathrsfs}
\usepackage[colorlinks, breaklinks, urlcolor={blue}, linkcolor={blue}, citecolor={blue}]{hyperref}
\usepackage{array}
\usepackage{amsmath}
\usepackage{mathrsfs}
\usepackage{type1cm}
\usepackage[english]{babel}
\usepackage{lmodern}
\usepackage{microtype}
\usepackage{booktabs}
\usepackage{caption}
\usepackage{braket}
\usepackage{xcolor}
\usepackage{orcidlink}

\begin{document}
\title{Environment engineering to protect quantum coherence in tripartite systems under dephasing noise}

\author{Sovik Roy \orcidlink{0000-0003-4334-341X} }
\email[]{s.roy2.tmsl@ticollege.org}
\affiliation{Department of Mathematics, Techno Main Salt Lake (Engg. colg.), \\Techno India Group, EM 4/1, Sector V, Salt Lake, Kolkata  700091, India}

\author{Aahaman Kalaiselvan}
\email[]{kaahaman.mct2022@citchennai.net}
\affiliation{Department of Mechatronics Engineering, Chennai Institute of Technology,\\ Chennai 600069, India}

\author{Chandrashekar Radhakrishnan\orcidlink{0000-0001-9721-1741}}
\email[]{chandrashekar.radhakrishnan@nyu.edu}
\affiliation{Department of Ccomputer Science and Engineering, NYU Shanghai\\, 567 West Yangsi Road, Pudong,
Shanghai, 200126, China.}

\author{Md Manirul Ali\orcidlink{0000-0002-5076-7619}}
\email[]{manirul@citchennai.net}
\affiliation{Centre for Quantum Science and Technology, Chennai Institute of Technology,\\ Chennai 600069, India}

\begin{abstract}
\noindent The practical success of quantum technology hinges on sustaining quantum coherence, which is vulnerable
to environmental interactions causing decoherence. We investigate coherence in tripartite quantum systems under the
influence of noisy environment. In this study, we explore the dynamics of the relative entropy of coherence for tripartite
pure and mixed states in the presence of structured dephasing environments at finite temperatures. Our findings
demonstrate that the system's resilience to decoherence is strongly influenced by the bath type and configuration.
Specifically, when each qubit interacts with an independent environment, the coherence dynamics differ from those
observed in a shared bath setting. In a Markov, memoryless environment, coherence in both pure and mixed states
decay faster, whereas coherence is preserved for longer time in the presence of environment memory. This highlights
the crucial role of environment memory in enhancing the robustness of tripartite coherence.
\end{abstract}


\maketitle

\section{Introduction}\label{sec:intro}

\noindent One of the important features of quantum mechanics is the quantum coherence which arises from the superposition principle \cite{Glauber:1963tx}. Quantum coherence is a resource and is central to the investigation of both entanglement and
other types of quantum correlations found in multipartite quantum states \cite{Horodecki:2009zz}. Quantum coherence in
multipartite systems has always been the center of study, as this significant feature of quantum states lies in the epicenter of
exploring these quantum states in doing quantum information processing tasks \cite{bennett2881,bennett1895,gisin145,Hillery:1998yq,Scarani:2005zz,nielsen2010}. The practical success of quantum
information processing depends on the sustained coherence of quantum states. However, interactions between quantum systems
and their environment induce decoherence, leading to the loss of quantum properties such as entanglement and coherence
\cite{zurek715,breuer2002}. A quantum system that interacts with the external noisy environment which is also modeled
quantum mechanically is known as an open quantum system. The dynamics of such a system is often described by differential
equation that governs the time evolution of the reduced density operator of the system, known as quantum master equation \cite{zurek715,scully97,breuer2016,PRXQ2024}. Approximations are often employed in deriving the master equation to
obtain the system's reduced density matrix. The Born approximation assumes weak coupling between the system and environment,
enabling a perturbative approach, while the Markov approximation assumes a short environment correlation time, justifying a
$\delta$-correlated environment correlation function \cite{scully97,breuer2002}.\\\\ However, many quantum devices exhibit non-Markovian (memory) effects, making the Markov approximation invalid. Such non-Markov behaviors are
particularly significant across various domains, including photonic and solid-state qubits, hybrid systems, quantum biology,
and quantum transport \cite{breuer2016,vega2017}. The environment memory is typically significant for structured environments,
where one can engineer the coupling between the system and the environment and use the environment memory to enhance the
longevity of quantum resources \cite{Braun2002,Sarlette2011,Nokkala2016,mazzola2009}.
In structured environments, the non-Markov dynamics
\cite{mazzola2009,bellomo07,wang2008,dajka2008,paz2008,junli2010,ali2010} of quantum systems exhibit distinct
characteristics compared to the memoryless behavior of Markov dynamics. The investigation of entanglement dynamics
in multipartite quantum systems has been an active area of research in recent years
\cite{sanma2007,aolita2008,lopez2008,weinstein2009,ankim2010,weinstein2010,christo2014,refx,refy,refz,reft}.
In this work, we employ relative entropy to investigate the multipartite coherence dynamics of quantum systems, considering
both pure and mixed states in the presence of a structured dephasing environment \cite{goan2010non,haikka2013non,guarnieri2014}. We consider different types of tripartite pure and mixed states. Since the considered tripartite states have different structures, the associated ability to maintain coherence in each state in response to local and common dephasing may be different. This motivates us to study different states to find out the best situation for coherence sustainability under the assumed dephasing and bath parameterization.
Two scenarios are explored: (a) individual qubit-environment interactions and (b) collective qubit interactions with a
common environment. The sustainability of tripartite coherence is evaluated under four conditions: (i) local Markov,
(ii) local non-Markov, (iii) common Markov, and (iv) common non-Markov dephasing environments.
Understanding coherence dynamics is crucial for quantum device fabrication, as it determines the operational time-frame
within which quantum operations must be completed before quantum resources
degrade \cite{shnirman2003noise,astafiev2006temperature,kakuyanagi2007dephasing}.\\\\ There are various measures to
quantify coherence, such as $l_{1}-$ norm, relative entropy of coherence, coherence measures
based on Jensen Shannon divergence. Baumgratz et al. introduced a method for measuring coherence based on a quantum resource-theoretic framework in Ref.~\cite{Baumgratz:2013ecx}. This resulted in several fundamental advancements in the field of resource theory
of quantum coherence \cite{brandao2015reversible,winter2016operational,chitambar2015relating,chandra2016,Streltsov:2016iow,
chitambar2019quantum}. Recently, the dynamics of quantum coherence have been explored in open quantum systems
\cite{ming2016,Pati2018Banerjee,chandra2019time,radhakrishnan2019dynamics,cao2020fragility,Ali:2022ajw,Ali:2022ntx,chengpra:2024}.
To quantify coherence we have considered the well-known measure of \textit{relative entropy of coherence}.
For any quantum state $\rho$, defined on the Hilbert space $\mathcal{H}$, the relative entropy of coherence is defined as \cite{Baumgratz:2013ecx,Streltsov:2016iow}
\begin{eqnarray}
\label{ref}
C_{R}(\rho) = \min_{\sigma \in \mathcal{I}} S(\rho \| \sigma).
\end{eqnarray}
The minimum is taken over the set of the incoherent states $\mathcal{I}$ which are states without quantum coherence.
Here $S(\rho \| \sigma) = {\rm Tr} (\rho \ln \rho - \rho \ln \sigma)$. In \cite{Baumgratz:2013ecx} it was proved that the
expression for the relative entropy also reduces to the form
\begin{eqnarray}
\label{ref1}
C_{R}(\rho) = S(\rho_{d}) - S(\rho),
\end{eqnarray}
where $\rho_{d}$ is the dephased state in the reference basis $\lbrace \vert i\rangle\rbrace$, i.e. the state obtained by deleting all off-diagonal entries. This measure enjoys the properties of physical interpretation and is easily computable. $S(\rho)$ is the von Neumann entropy of
the state $\rho$. \\\\The paper is organized as follows. In section $II$, we have described the open quantum system where, model Hamiltonian of three qubits individually subjected to local dephasing environment and three qubits exposed to common dephasing environment together, have been discussed. A schematic diagram has also been shown in this section. Section $III$ elaborates results and discussion part where the three qubit pure and mixed states have been subjected to Markov (local and common baths) and non-Markov (local and common baths) dephasing noise. This is followed by conclusion and future direction in section $IV$.

\section{Description of the Models}\label{sec:Model}
\noindent We now discuss the models of three-qubits system subjected to two kinds of noisy environments, viz. (i) local dephasing and (ii) common dephasing as shown in Fig.~\ref{schematic}.

\subsection{Three Qubits Under Local Dephasing Environment}\label{subsec:Local}
\noindent
We examine a spin-boson model comprising three non-interacting qubits, each interacting with a local bosonic environment,
as depicted in Fig.~\ref{schematic}a. The combined Hamiltonian for the three qubits and their environment is described by
\begin{eqnarray}
\label{totalhamiltonianldp}
H = \sum_{i=1}^3\Big[\frac{\hbar}{2}\omega_{0}^i\sigma_{z}^i + \sum_{k}\hbar \omega_{ik}b_{ik}^{\dagger}b_{ik} + \sigma_{z}^i
(B_{i} + B_{i}^{\dagger})\Big].
\end{eqnarray}
In this model, $B_{i}=\hbar \sum_{k}g_{ik}b_{ik}$, with $\omega_{0}^i$ and $\sigma_{z}^i$ representing the transition frequency and Pauli spin operator of the $i^{th}$ qubit, respectively. We assume that all three qubits have the same transition frequency $\omega_{0}$. Each local environment of the $i^{th}$ qubit is modeled as an ensemble of infinite bosonic modes with frequencies $\omega_{ik}$, where $b_{ik}^{\dagger}$ and $b_{ik}$ are the creation and annihilation operators for the $k^{th}$ mode of the local environment interacting with the $i^{th}$ qubit. The coupling strength between the $i^{th}$ qubit and its local environment is represented by $g_{ik}$. In the continuum limit, $\sum_k |g_{ik}|^2 \rightarrow \int d\omega J_{i}(\omega) \delta(\omega_{ik}-\omega)$, where $J_{i}(\omega)$ denotes the spectral density of the local environment for the $i^{th}$ qubit. Initially, we consider the three-qubit system to be decoupled from their respective environments, with each local environment initially in thermal equilibrium at temperature $T_i$. As the system evolves under the Hamiltonian $H$, we trace out the environment degrees of freedom to obtain the reduced density matrix for the quantum system, which allows us to study its dynamics. The decay dynamics of the reduced density matrix $\rho(t)$ for this three-qubit system under local dephasing is described by the quantum master equation
\begin{eqnarray}
\label{NL}
\frac{d}{dt} \rho(t) = -\frac{i}{\hbar} \big[ H_S, \rho(t) \big] +
\sum_{i=1}^{3} \gamma_{i}(t)  \Big( \sigma_z^i \rho(t) \sigma_z^i - \rho(t) \Big),
\end{eqnarray}
where
\begin{eqnarray}
\label{Hs}
H_S=\frac{\hbar}{2} \sum_{i=1}^3 \omega_0 \sigma_{z}^i
\end{eqnarray}
is the system Hamiltonian representing three qubits and consequently the dephasing rate is given by
\begin{eqnarray}
\label{gammat}
\gamma_{i}(t) = 2 \int_{0}^{\infty} d\omega  J_{i}(\omega) \coth\left(\frac{\hbar \omega}{2 k_B T_i}\right) \frac{\sin(\omega t)}{\omega}.
\end{eqnarray}
The time-dependent dephasing rate $\gamma_{i}(t)$ is determined by the spectral density $J_{i}(\omega)$.
For this model, we consider an Ohmic-type spectral density given by \cite{Leggett:1987zz}
\begin{eqnarray}
J_{i}(\omega) = \eta_{i} \omega \exp\left(-\frac{\omega}{\Lambda_i} \right).
\label{ohm}
\end{eqnarray}
Typically, the environment is assumed to be large enough to quickly reset to its initial state. However, we investigate the environment memory effect on the coherence dynamics of three-qubit states with a finite cut-off frequency $\Lambda_i$. Under the Markov approximation, where the environment's correlation time is much shorter than the system dynamics' timescale, the time-dependent coefficient $\gamma_{i}(t)$ can be replaced by its long-time Markov value $\gamma_{i}^M=4 \pi \eta_i k_B T_i/\hbar$. In this scenario, the decay dynamics of the density matrix $\rho(t)$ is defined as Markov. The decay constant $\gamma_{i}^M$ simplifies to $\gamma_0=4 \pi \eta k_B T/\hbar$ assuming uniform spectral densities with uniform coupling strengths $\eta_i=\eta$, and equal temperatures $T_i=T$ for each local environment.
The Markovian assumption holds when the bath's relaxation time is much shorter than the system's evolution time. However,
if the bath relaxation time is comparable to the system's evolution time, the environment's memory effects become significant.
In this case, the decoherence dynamics of the three qubits is governed by the master equation (\ref{NL}), with the time-dependent
dephasing rate $\gamma(t)$ given by equation (\ref{gammat}).

\subsection{Three Qubits Under Common Dephasing Environment}\label{subsec:common}
\noindent
Next, we examine a scenario where the three qubits interact with a common environment as shown in Fig.~\ref{schematic}b.
The microscopic Hamiltonian for the three two-level systems coupled to a common environment is expressed as
\begin{eqnarray}
\label{Dephc}
H = \frac{\hbar}{2} \sum_{i=1}^{3}\Big[ \omega_0^i \sigma_z^i + \sum_k \hbar \omega_{k} b_{k}^{\dagger} b_{k}+ S_z \left( B + B^{\dagger} \right)\Big].
\end{eqnarray}
Here, $S_z = \sum_i \sigma_z^i$ represents the collective spin operator for the three-qubit system, while the environment operator is given by $B=\hbar \sum_k g_{k} b_{k}$. It is assumed that all three qubits have the same transition frequency, $\omega_0^i = \omega_0$. The common environment is modeled as a collection of bosonic field modes with frequencies $\omega_{k}$, where $b_{k}$ and $b_{k}^{\dagger}$ are the annihilation and creation operators associated with the $k^{th}$ mode of the environment. We start with a factorized initial state where the system and environment are decoupled, and the environment is initially in thermal equilibrium at some temperature $T$. The quantum master equation for the three qubits interacting with this common environment at finite temperature is given below.
\begin{eqnarray}
\frac{d}{dt} \rho(t) = -\frac{i}{\hbar} \big[ H_S, \rho(t) \big] +
\gamma(t) S_z \rho(t) S_z -\alpha(t) S_z S_z \rho(t) - \alpha^{\ast}(t) \rho(t) S_z S_z,
\label{Nc}
\end{eqnarray}
where $\gamma(t)$ is given by equation (\ref{gammat}) and
\begin{eqnarray}
\alpha(t) = \int_{0}^{\infty} d\omega  J(\omega) \coth\left(\frac{\hbar \omega}{2 k_B T}\right) \frac{\sin(\omega t)}{\omega}
- i \int_{0}^{\infty} d\omega  J(\omega) \frac{1-\cos(\omega t)}{\omega}.
\label{alphat}
\end{eqnarray}
The quantum master equation simplifies under the Markov approximation with negligible environment memory as follows;
\begin{eqnarray}
\frac{d}{dt} \rho(t) = -\frac{i}{\hbar} \big[ H_S, \rho(t) \big] + \frac{\gamma_0}{2}  \Big( 2 S_z \rho(t) S_z  - S_z S_z \rho(t)
-  \rho(t) S_z S_z \Big).
\label{Mc}
\end{eqnarray}
Next, by considering specific three-qubit states, we demonstrate how environment memory effects can enhance the robustness of multi-qubit coherence in a common environment. We analyze the coherence dynamics under both Markov and non-Markov dephasing for an Ohmic spectral density $J(\omega)=\eta \omega \exp (-\omega/\Lambda)$. In our numerical data, we choose $k_B T=\hbar \omega_0/4\pi$,
resulting in $\gamma_0=\eta \omega_0$. The other parameter values are set as $\eta=0.1$ and $\Lambda=10^{-2}\omega_0$.\\\\
The plot in Fig.~\ref{schematic} shows schematic diagram where each qubit of the tripartite state is subjected to local noisy environment (Fig.~\ref{schematic}a) as well as to common noisy environment (Fig.~\ref{schematic}b).
\begin{figure}[h]
\label{schematic}
\includegraphics[width=12.12cm]{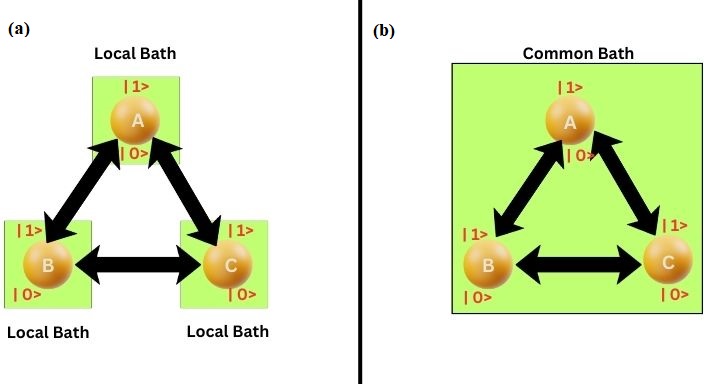}
\caption{The figure shows schematic representation of the situation where the qubits are subjected to local (left) and common (right) bath. This local or common bath may be Markov (memoryless) or non-Markov.}
\label{schematic}
\end{figure}
In the discussion that follows, we have considered pure and mixed tripartite states that are subjected to dephasing setting. Here the qubits are distributed among parties Alice (A), Bob (B) and Charlie (C). Each qubit is exposed individually to local Markov and non-Markov baths or all the qubits are subjected to common Markov and non-Markov baths.\\\\
The quantum master equations given through Eq. (\ref{NL}) and  Eq. (\ref{Nc}), along with the time-dependent coefficients $\gamma_{i}(t)$ in Eq. (\ref{gammat}) and $\alpha(t)$ in Eq. (\ref{alphat}), are derived using second-order perturbation theory. We assume a weak system-environment coupling to justify this approach \cite{breu2002,car1999}.

\section{Pure States subjected to Markov and non-Markov Dephasing Noise}

\noindent In this section, we explore the relative entropy of coherence in several tripartite pure states exposed to dephasing environment. For our investigation, we have considered tripartite pure states such as GHZ, W, $W\overline{W}$ and Star. Before delving into our analysis, we will first outline the significance of these tripartite states. The coherence dynamics of these tripartite states is studied under local and common dephasing environment using relative entropy of coherence. The tripartite states are categorized into classes namely-the GHZ class and W class, based on Stochastic Local Operations and Classical Communication (SLOCC). Among the two aforementioned categories of tripartite states, the GHZ state represent a truly tripartite entangled state, so if one of its qubit is lost, the entire entanglement is destroyed. In contrast, the W state lacks genuine tripartite entanglement, instead its entanglement shared in a bipartite manner. As a result, when any qubit in a W state undergoes decoherence, the state retains some degree of entanglement. The states GHZ and W are defined as \cite{Greenberger:1989tfe,Dur:2000zz}
\begin{eqnarray}
    \vert GHZ\rangle &=& \frac{1}{\sqrt{2}}\Big(\vert 000\rangle + \vert 111\rangle\Big)\nonumber\\
    \vert W\rangle &=& \frac{1}{\sqrt{3}}\Big(\vert 100\rangle + \vert 010\rangle + \vert 001\rangle\Big).
    \label{ghzwpure}
\end{eqnarray}
Further we investigate two other highly significant tripartite pure states viz. the $\vert W\overline{W}\rangle$ and $\vert Star\rangle$ state \cite{Roy:2022qoy,radhakrishnan1}. The $W\overline{W}$ state is an equal superposition of W-state and $\overline{W}$-state,
defined as
\begin{eqnarray}
\label{wwbar}
\vert W\overline{W}\rangle = \frac{1}{\sqrt{2}}\Big(\vert W\rangle + \vert \overline{W}\rangle\Big),
\end{eqnarray}
where the state $\vert \overline{W} \rangle$ is the spin flipped version of $\vert W \rangle$ given by
\begin{eqnarray}
\vert \overline{W}\rangle = \frac{1}{\sqrt{3}}\Big(\vert 011\rangle + \vert 101\rangle + \vert 110\rangle\Big).
\end{eqnarray}
The type of state $W\overline{W}$ is chosen because it has quantum coherence at the single-qubit,bipartite, and tripartite levels, as
well as bipartite and tripartite quantum correlations. Such a state is a good test bed for studying quantum correlations distributed at
different levels. However, The GHZ, $W\overline{W}$ and W-states are symmetric states and entanglement in their reduced
bipartite form does not depend on which qubit has been subjected to decoherence effects. In contrast to these states is the Star
state, which is asymmetric.
\begin{eqnarray}
\label{star}
\vert Star\rangle = \frac{1}{2}\Big(\vert 000\rangle + \vert 100\rangle + \vert 101\rangle + \vert 111\rangle\Big).
\end{eqnarray}
Like the $W\overline{W}$  state, the Star state also has coherence and correlations distributed at all possible levels.
The Star state exhibits asymmetry due to its uneven distribution of correlations. It consists of two types of qubits, viz
(i) peripheral qubits and
(ii) a central qubit.
The first two qubits are peripheral and the third is central qubit in the Star state. If the central qubit is traced out, the other qubits become separable. Conversely, if we perform partial trace over the first and second qubits, the remaining qubits still retain some entanglement. We shall consider now these four types of tripartite pure states defined in eqs.(\ref{ghzwpure}-\ref{star}) and discuss the change in the coherence of the states when they are subjected to dephasing dynamics.\\\\

\subsection*{Relative entropy of coherence in pure states:}
\begin{figure}[h]
\label{plotpures}
\includegraphics[width=15.0cm]{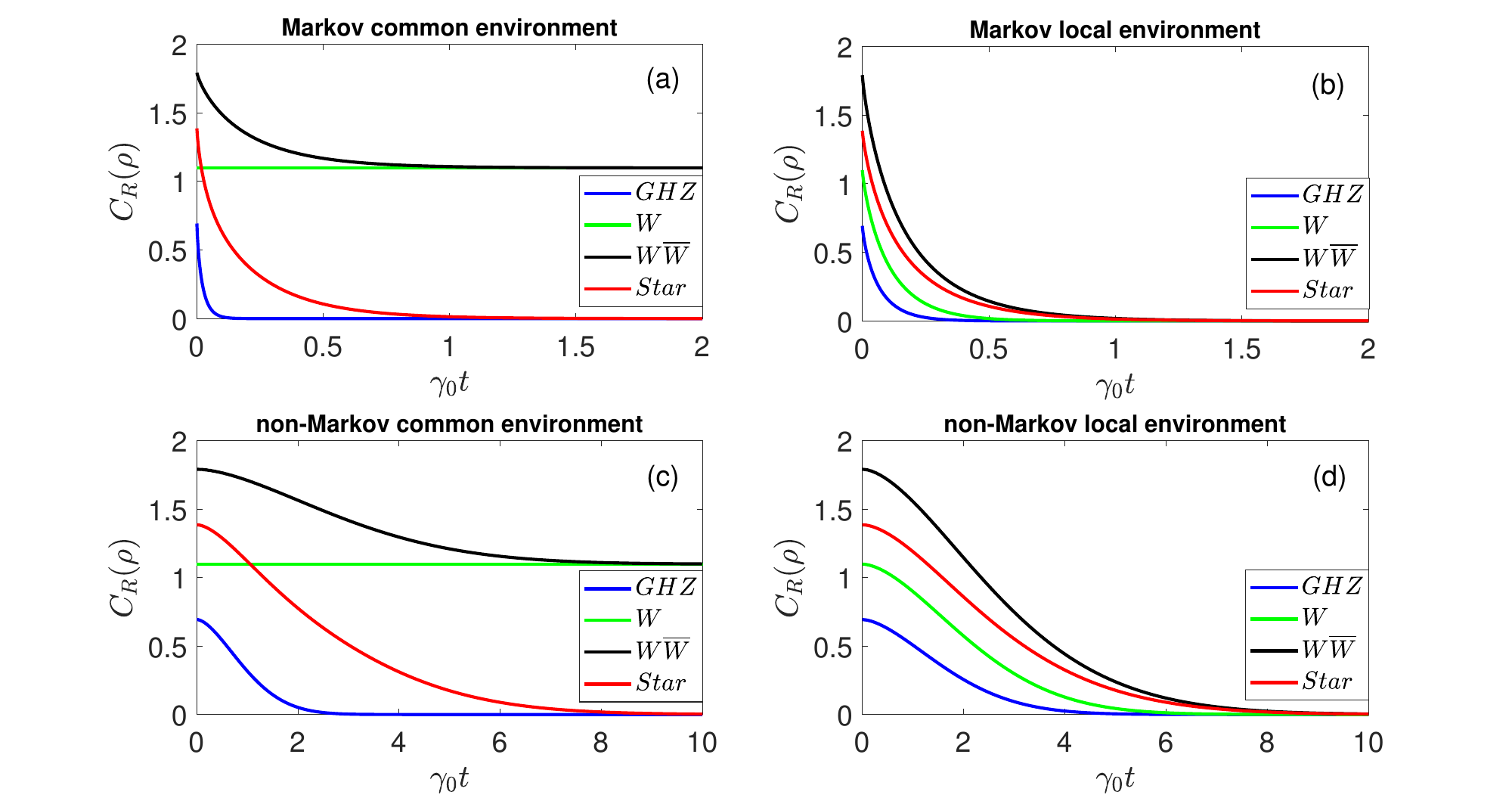}
\caption{Here fig (a) denotes dynamics of coherence of all four pure states in Markov common environment and fig (b) denotes the dynamics of coherence of the pure states in Markov local environment while figs. (c) and (d) respectively denote the coherence dynamics of these pure states in non-Markov common and local environments.}
\label{plotpures}
\end{figure}
\noindent In the Fig.~\ref{plotpures}, we show the dynamics of coherence for pure tripartite states under Markov (common and local)
and non-Markov (common and local) environments. This study aims to provide clearer insights into how the relative entropy of
coherence, denoted as $C_{R}(\rho)$, varies for the pure states $\vert GHZ\rangle$, $\vert W\rangle$, $\vert W\overline{W}\rangle$
and $\vert Star\rangle$ under different types and configurations of dephasing environments.

\subsubsection*{Markov environment:}

\noindent \textit{Common bath:} In Markov common environment i.e. (Fig.~\ref{plotpures}a) we see that, relative entropy of coherence of
W-state {\it i.e.} $C_{R}(W)$ remains constant and does not decay over time, while the relative entropy of coherences of GHZ, $W\overline{W}$ and Star states decay over time. The amount of coherence of the initial states at time $t=0$ are as follows
$C_{R}(W\overline{W})>C_{R}(Star)>C_{R}(W)>C_{R}(GHZ)$. However the decay rates of these three pure states {\it i.e.}
$W\overline{W}$, Star, and GHZ-state differ. The rate of decay of relative entropy of coherence of GHZ state is
much higher than the rest. The coherence of GHZ-state $C_{R}(\rho_{GHZ})$ vanishes at $\gamma_{0}t \sim 0.1$,
whereas $C_{R}(\rho_{Star})$ completely decays when $\gamma_{0}t \sim 1.0$. The coherence $C_{R}(\rho_{W\overline{W}})$
of $W\overline{W}$ decays over time but eventually saturates at a value of $1$ in the long-time limit.
\noindent \textit{Local bath:} Fig.~\ref{plotpures}b illustrates the coherence dynamics of the pure states GHZ, W,
$W\overline{W}$, and Star under Markov local environment. All these pure states exhibit exponential decay of
coherence under Markovian local
environment. However, a key difference emerges between Markovian common and local environments concerning the W-state: while the W-state remains decoherence-free in a Markovian common environment, it undergoes exponential decay in a Markovian local environment
Also, a slow decay of coherence in $W\overline{W}$ is seen, when subjected to Markov common bath, whereas an enhanced decay
happens for this state under Markov local environment. A complete decay of coherence for the state $W\overline{W}$ occurs under
Markov local environment, unlike the case of Markov common environment where $C_{R}(\rho_{W\overline{W}})$ saturates to a finite
value in the long-time limit.

\subsubsection*{non-Markov environment:}

\noindent The time variation of coherence of the tripartite pure states in a common non-Markov dephasing environment is depicted
in Fig.~\ref{plotpures}c. Under non-Markov common environment (see Fig.~\ref{plotpures}c), W-state remains
decoherence-free. Comparing the decoherence dynamics of the GHZ, $W\overline{W}$, and Star states under Markov
(Fig.~\ref{plotpures}a) and non-Markov (Fig.~\ref{plotpures}c) common environments reveals that coherence decay is
significantly slower in the non-Markovian case for all three states.
The non-Markovian coherence dynamics of the tripartite pure states under local dephasing environments is
shown in Fig.~\ref{plotpures}d. Next, we compare the coherence dynamics of all tripartite pure states under Markov local (Fig.~\ref{plotpures}b) and non-Markov local (Fig.~\ref{plotpures}d) environments. Slower decay of coherence is
observed for all the states under non-Markov local environments, compared to Markov local case. Hence, we see
that the coherence decay of a non-Markov process is slower than that of a Markov process. Interestingly, unlike in the
non-Markovian common environment, the W-state is not decoherence-free in a non-Markovian local environment.
Overall, the sustainability of quantum coherence is significantly improved in the presence of environment memory when
qubits interact with structured environments.\\\\
We observed that the W-state in the common bath environment lies in the decoherence-free subspace (DFS), which can easily be identified as follows. In our common dephasing model (\ref{Dephc}), the interaction Hamiltonian for the tripartite system coupled to a common environment is given by the term $H_I=S_z( B +B^{\dagger})$. Since the W-state is an eigenstate of the collective spin operator $S_z$ for the three-qubit system, it remains in the DFS \cite{lidar19994556} under this system-environment interaction, as evidenced by the green lines in Figures \ref{plotpures}a and \ref{plotpures}c.
\noindent
\subsection{Mixed states subjected to Markov and non-Markov dephasing environment}
\noindent Due to practical difficulties in preparing and preserving pure states under standard conditions, focussing on coherence in mixed states become necessary for a comprehensive understanding. To this purpose, we examine the following mixed states such as (i) mixtures of GHZ and W \cite{lohmayer2006}, (ii) Werner-GHZ \cite{acinref} and (iii) Werner-W \cite{park2009,lohmayer2006}.

\subsubsection*{Mixture of GHZ and W state}
\noindent The mixture of GHZ and W state is defined as
\begin{eqnarray}
\label{ghzw1}
\rho_{GW} &=& p\vert GHZ\rangle\langle GHZ\vert + (1-p)\vert W\rangle\langle W\vert.
\end{eqnarray}
where $p$ is the probability of mixing.
\begin{figure}[h]
\includegraphics[width=15.0cm]{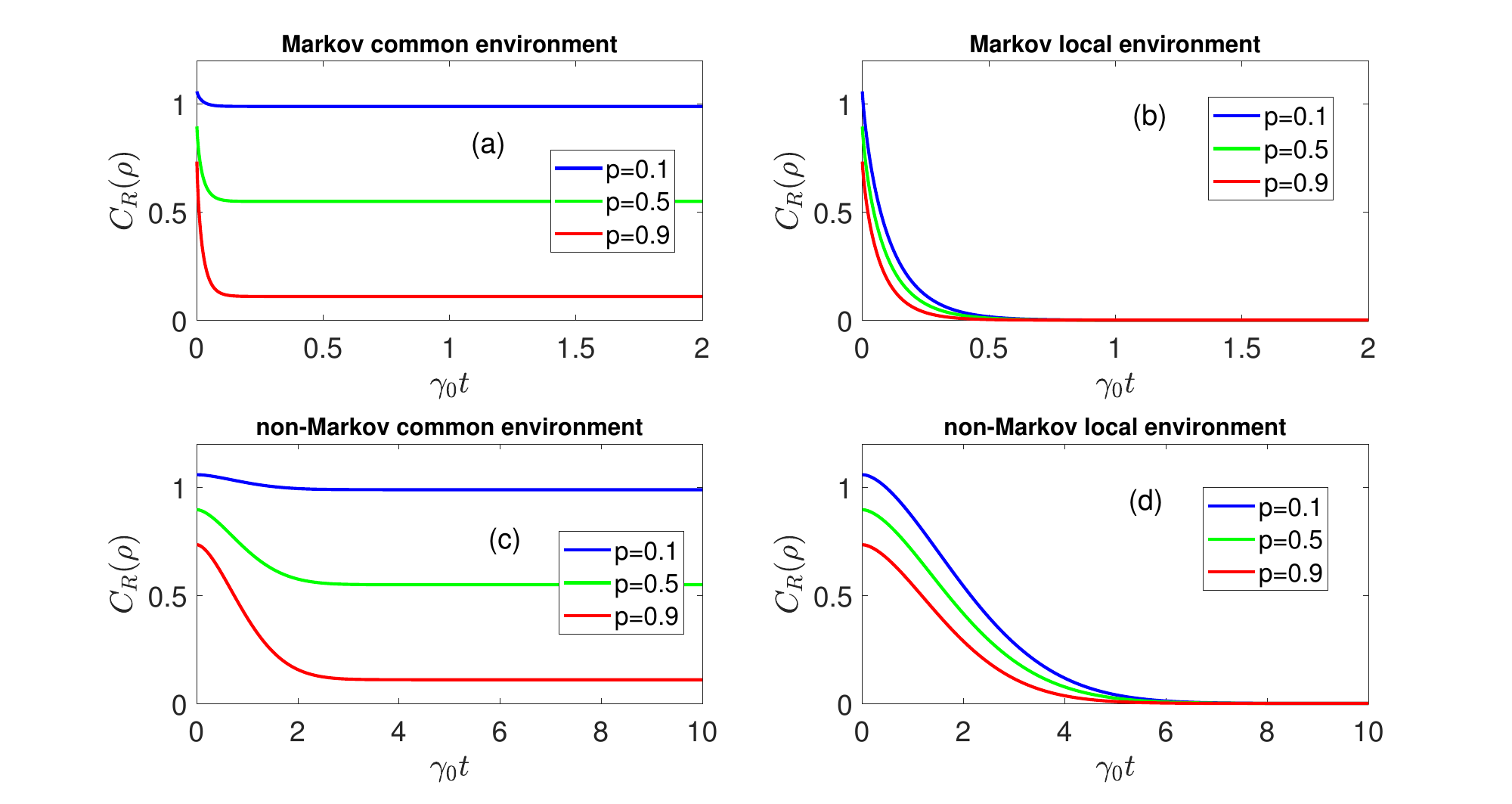}
\caption{Here $(a$) represents the relative entropy of coherence $C_{R}(\rho)$ of $\rho_{GW}$ in Markov common
environment, $(b)$ shows $C_{R}(\rho)$ of $\rho_{GW}$ in Markov local environment while Figs.~$(c)$ and $(d)$ are
showing dynamics of $C_{R}(\rho)$ of $\rho_{GW}$ in non-Markov common and local environments.}
\label{ghzw}
\end{figure}
In Figure~\ref{ghzw}, we present the dynamics of coherence $C_{R}(\rho_{GW})$ of the mixed states $\rho_{GW}$
for different mixing parameters $p$. Three different cases $p=0.1$ (blue), $p=0.5$ (green), and $p=0.9$ (red) are
examined. The robustness of coherence is analyzed under Markovian and non-Markovian dephasing environments,
both common and local.
\vskip 0.5cm
\noindent \textit{Common bath:} Fig.~\ref{ghzw}a shows the coherence dynamics of the state $\rho_{GW}$ under Markovian
common environment. Under Markov common dephasing, the coherence of the state $\rho_{GW}$ remains nearly
unchanged over time for the mixing parameter $p=0.1$, whereas for $p=0.5$ and $p=0.9$, decoherence is observed
for a very short duration. At $\gamma_{0}t=0$, the initial coherence values of the states are as follows:
$1<C_{R}(\rho_{GW})<1.1$ for $p=0.1$, $0.7<C_{R}(\rho_{GW})<0.8$ for $p=0.9$, and the initial value of
$C_{R}(\rho_{GW})$ is slightly below $0.9$ when $p=0.5$. Under Markov common dephasing (Fig.~\ref{ghzw}a),
the coherence of all three states rapidly converges to distinct steady-state values. The non-Markovian decoherence
dynamics of the mixed states $\rho_{GW}$ under a common dephasing environment is illustrated in Fig.~\ref{ghzw}c.
Under non-Markov common environment, we do not see a rapid decay of coherence. Instead, the coherence decreases
slowly and approaches steady saturation values.
\vskip 0.5cm
\noindent \textit{Local bath:} In Fig.~\ref{ghzw}b, we show the coherence dynamics of the mixed states $\rho_{GW}$ under
Markov local environment. The states $\rho_{GW}$ with mixing parameters $p=0.1,~0.5,~0.9$ exhibit exponential
decay of coherence under Markovian local dephasing. We notice that under a Markovian local environment, $\rho_{GW}$
experiences complete coherence loss within a short interval ($\gamma_{0}t\sim0.5$). In contrast, under a Markovian
common environment, the same mixed states exhibit long-time saturation value of coherence. Figure~\ref{ghzw}d illustrates
the coherence dynamics of the three mixed states in a non-Markovian local bath. Next, we compare the decoherence dynamics of
the states $\rho_{GW}$ under Markov and non-Markov local environments. For local noise, we see that the longevity of coherence
increases under non-Markov dephasing (Figure~\ref{ghzw}d) for all three states. Hence, non-Markovian dynamics is beneficial
for preserving quantum coherence for a longer duration compared to memoryless Markovian dynamics.

\subsubsection*{Werner-GHZ state}
\noindent The Werner-GHZ is a mixed state which is defined as
\begin{eqnarray}
\label{ghzwerner}
\rho_{GR} = p\vert GHZ\rangle\langle GHZ\vert + \frac{(1-p)}{2} \mathbb{I}_{8},
\end{eqnarray}
where as before $p$ is the probability of mixing and $ \mathbb{I}_{8}$ is the identity matrix of order $8$. The state in Eq. (\ref{ghzwerner}) is a regular tripartite state which is a mixture of GHZ and identity $\mathbb{I}_{8}$.
\begin{figure}[]
\label{wernerghz}
\includegraphics[width=15.0cm]{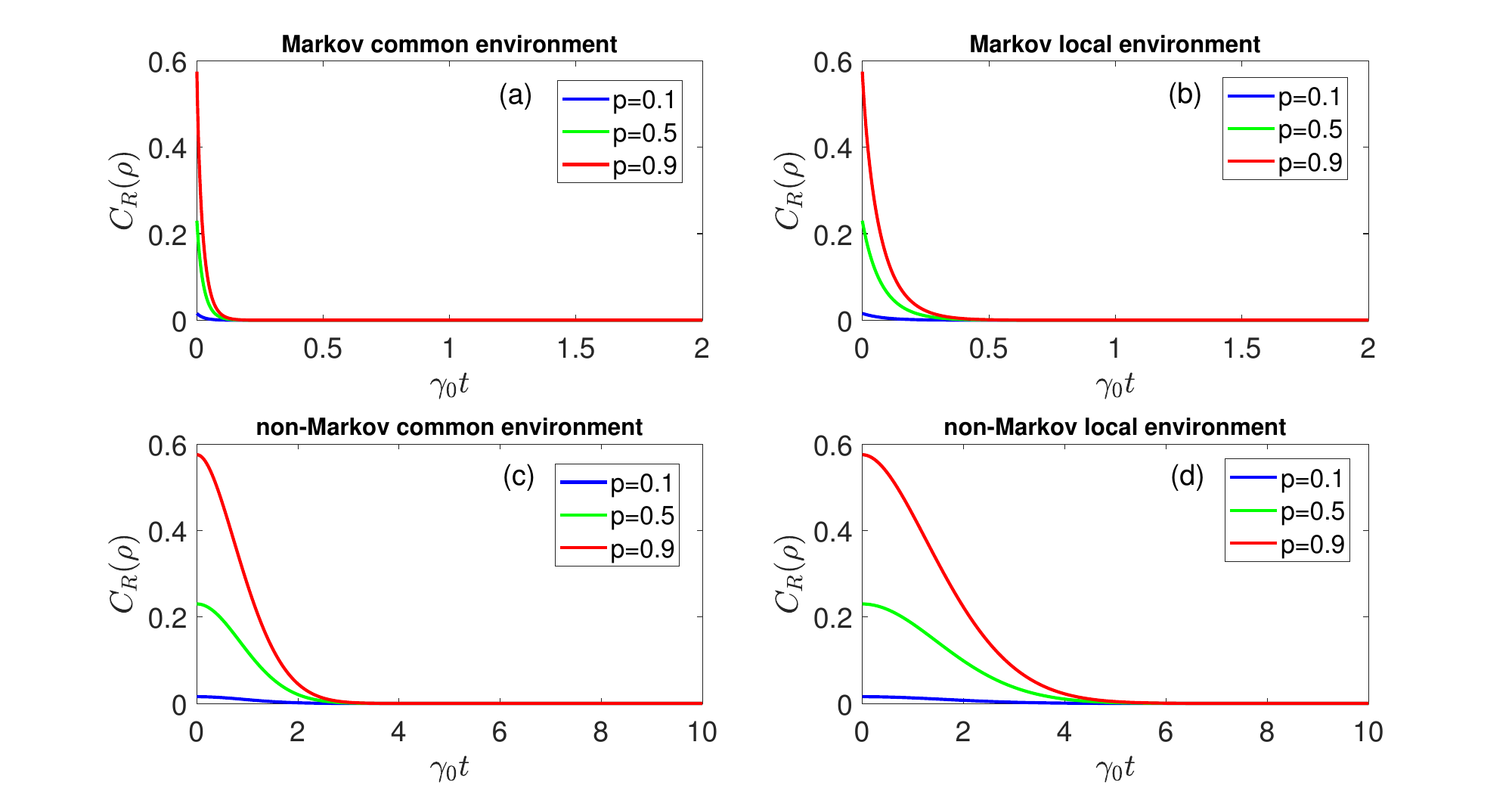}
\caption{Here Fig.~$(a)$ represents the relative entropy of coherence $C_{R}(\rho)$ of $\rho_{GR}$ in Markov common environment, fig. $(b)$ shows $C_{R}(\rho)$ of $\rho_{GR}$ in Markov local environment while Figs.~$(c)$ and $(d)$ are showing dynamics of $C_{R}(\rho)$ of $\rho_{GR}$ in non-Markov common and local environments.}
\label{wernerghz}
\end{figure}

\vskip 0.5cm

\noindent \textit{Common bath:} In Fig.~\ref{wernerghz}, we present the dynamical evolution of quantum coherence for the
state described in Eq.~(\ref{ghzwerner}) as a function of $\gamma_{0}t$. At $\gamma_{0}t=0$, initial coherence
follows the hierarchy $C_{R}(\rho_{GR})_{p = 0.9}>C_{R}(\rho_{GR})_{p = 0.5}>C_{R}(\rho_{GR})_{p = 0.1}$.
From Fig.~\ref{wernerghz}a, it is evident that, under a Markovian common bath, the coherence measure $C_{R}(\rho_{GR})$
exhibits exponential decay for mixing probabilities $p=0.5$ and $p=0.9$. For probability of mixture $p=0.1$, almost there is
no coherence observed, slight coherence is seen over a negligible time interval. The decoherence dynamics of the initial
Werner-GHZ mixed states $\rho_{GR}$ under non-Markovian common environment is shown in Fig.~\ref{wernerghz}c.
For the initial tripartite state $\rho_{GR}$, coherence persists longer when the qubits interact with a common {\it non-Markov}
environment than with a common {\it Markov} environment.
\vskip 0.5cm
\noindent \textit{Local bath:} In Markov local environment (Fig.~\ref{wernerghz}b), there is exponential decay of coherence of the
state $\rho_{GR}$, for mixing parameters $p=0.5$ and $p=0.9$. Notably, the initial coherence for $p=0.9$ is greater than that
for $p=0.5$. The coherence decay of the state $\rho_{GR}$ is slower under Markov local noise compared to a Markov common
dephasing. Interestingly, the coherence of the state $\rho_{GR}$ is the most fragile with respect to Markov common environment.
For $p=0.1$, the state $\rho_{GR}$ exhibits negligible coherence that persists only briefly in a Markov local bath. In
Figure~\ref{wernerghz}d, we demonstrate the coherence dynamics of the tripartite mixed states $\rho_{GR}$ under non-Markovian
local environments. For the initial states $\rho_{GR}$, coherence decay is significantly slowed down under non-Markovian local
environments (Fig.~\ref{wernerghz}d) compared to the Markov local case. We again confirm that the environment memory can
enhance the robustness of coherence for multi-qubit systems.

\subsubsection*{Werner-W state}
\noindent The next mixed state that we consider here is Werner-W state, which is defined as
\begin{eqnarray}
\label{wernerw}
\rho_{W\!R} = p\vert W\rangle\langle W\vert + \frac{(1-p)}{2} \mathbb{I}_{8},
\label{wnws}
\end{eqnarray}
where $p$ is the probability of mixing and $ \mathbb{I}_{8}$ is the identity matrix of order $8$. The state in
Eq.~(\ref{wernerw}) is a regular tripartite state, a mixture of W-state and $\mathbb{I}_{8}$.
\begin{figure}[h]
\label{wrw}
\includegraphics[width=15.0cm]{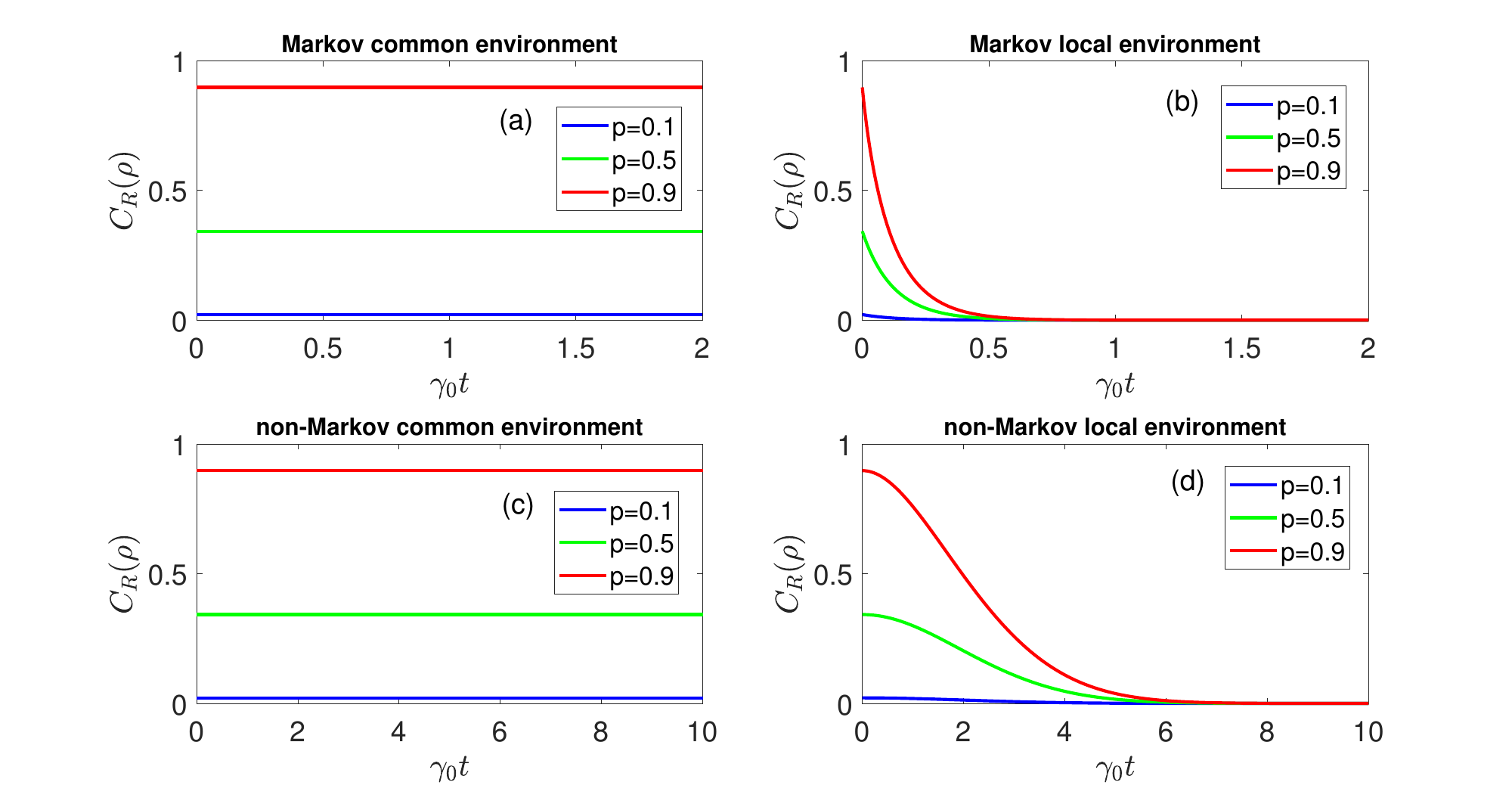}
\caption{Here Fig. (a) represents the relative entropy of coherence $C_{R}(\rho)$ of $\rho_{W\!R}$ in Markov common environment, fig. (b) shows $C_{R}(\rho)$ of $\rho_{W\!R}$ in Markov local environment while figs. (c) and (d) are showing dynamics of $C_{R}(\rho)$ of $\rho_{W\!R}$ in non-Markov common and local environments.}
\label{wrw}
\end{figure}
The coherence dynamics of the mixed Werner-W state (\ref{wernerw}) is shown in Fig.~\ref{wrw}. At $t=0$, the initial
coherence of $\rho_{W\!R}$ is nearly zero for a mixing parameter of $p=0.1$. For $p=0.5$, the initial coherence ranges
between $0.3$ and $0.4$, while for $p=0.9$, it exceeds $0.8$ but remains below $1$. The coherence dynamics of $\rho_{W\!R}$
is quite interesting. We see from Figs.~ \ref{wrw}(a) and \ref{wrw}(c) that $C_{R}(\rho_{W\!R})$ i.e. coherence of $\rho_{W\!R}$
does not decay in either Markov common or non-Markov common environment. This happens because the mixed
state $\rho_{W\!R}$ is invariant under both non-Markov and Markov dynamics in common
environment. The reason is as follows: we have already seen that the W-state remains decoherence free under
non-Markov and Markov common dephasing. We have also verified that the maximally mixed state $\mathbb{I}_{8}/8$
remains invariant under both Markov and non-Markov dynamics in common environment. In other words, the dynamical
maps corresponding to the master equations (\ref{Nc}) and (\ref{Mc}) are unital, preserving the identity operator.
Moreover, these dynamical maps are linear by construction. Consequently, when they act on the state
$\rho_{W\!R}$, they do so linearly on both components: the W-state projector $\vert W\rangle\langle W\vert$ and the
maximally mixed state $\mathbb{I}_{8}/8$. Since both $\vert W\rangle\langle W\vert$ and $\mathbb{I}_{8}/8$ are individually
invariant under the evolution governed by the master equations (\ref{Nc}) and (\ref{Mc}), it follows that any Werner-W
state of the form (\ref{wnws}) remains unaffected by decoherence in these scenarios. The amount of initial coherence in Werner-W
states depends on the mixing probability $p$. This coherence value is preserved when the states are in contact with a common bath.
From Fig.~\ref{wrw}b, we observe that the Markov local environment induces distinct dynamics for the mixed state $\rho_{W\!R}$
depending on the value of the mixing parameter $p$. When $p=0.1$, the coherence remains negligibly small throughout the entire
range of $\gamma_{0}t$. For $p=0.5$, the coherence $C_{R}(\rho_{W\!R})$ starts with a value greater than $0.3$, then decays
exponentially over time, eventually vanishing at approximately $\gamma_{0}t \sim 0.5$. When the mixing parameter is increased to
$p=0.9$, the coherence starts at a value greater than $0.8$ at $\gamma_{0}t=0$ and follows an exponential decay, ultimately
reaching zero and leading to a complete loss of coherence at around $\gamma_{0}t \sim 0.7$.
Fig.~\ref{wrw}d illustrates the coherence dynamics of the tripartite mixed states $\rho_{W\!R}$ in non-Markovian
local environments. We see that the coherence decay of $\rho_{W\!R}$ is significantly slower for mixing parameters
$p=0.5$ and $p=0.9$ under non-Markovian local environments (Fig.~\ref{wrw}d) compared to its Markovian local
counterpart (Fig.~\ref{wrw}b).

\section{conclusion:}

\noindent Environmental decoherence threatens quantum technology. We have explored tripartite coherence under noisy conditions, showing that bath configuration and memory significantly impact resilience. Shared baths and environment memory enhance coherence preservation, crucial for robust quantum systems. For our study we have applied relative entropy of coherence as the measure of coherence of the system. We have examined here a spin-boson model encompassing three non-interacting qubits system that have been subjected to noisy environments such as (a) local dephasing and (b) common dephasing, which have again been sub-categorized into (i) Markovian model and (ii) non-Markovian model for each of the local and common dephasing environment. For our study we have considered pure states such as GHZ, W, $W\overline{W}$ and Star while mixture of GHZ and W and mixed states such as Werner-GHZ and Werner-W have also been taken into consideration. In a Markov common environment, W-state coherence remains constant, while GHZ, Star, and $W\overline{W}$ coherence decays, with GHZ decaying fastest. Initial coherence ranking is $W\overline{W} >  Star > W > GHZ$. GHZ coherence vanishes rapidly, followed by Star, while $W\overline{W}$ saturates at a value above $1$. In a Markov local environment, all states exhibit exponential coherence decay. The W-state exhibits a key difference: stable coherence in a Markov common environment, versus exponential decay in a local one. Similarly, $W\overline{W}$ shows slow decay in a common bath but complete decay in a local bath, without saturating to a finite value. In non-Markov common environments, the W-state remains decoherence free, and GHZ, $W\overline{W}$ and Star states decay slower than in Markov cases. Non-Markov local environments also show slower decay than Markov local, though the W-state loses coherence. Thus, environment memory consistently enhances coherence preservation. Under Markov common noise, the coherence of the mixture $\rho_{GW}$ stabilizes, with minimal decay for $p=0.1$ and brief decay for $p=0.5,\:0.9$. Non-Markov common noise results in slower decay and steady saturation. In contrast, Markov local noise causes rapid, complete coherence loss for all $p$ values. Non-Markov local noise slows decay, extending coherence longevity. Thus, environment memory consistently enhances coherence preservation compared to Markovian dynamics. Markov local noise rapidly depletes the coherence of $\rho_{GW}$, unlike the saturation seen in common environments. Non-Markov local noise significantly extends coherence longevity compared to Markov local, demonstrating the benefit of environment memory. On the other hand, For the Werner-GHZ (i.e. $\rho_{GR}$) state, initial coherence decreases with decreasing mixing probability. Under Markov common noise, coherence decays exponentially for higher mixing probabilities, with minimal coherence for $p=0.1$. Non-Markov common noise extends coherence duration compared to Markov. Markov local noise causes exponential $\rho_{GR}$ coherence decay, slower than common noise, with $p=0.1$ showing minimal coherence. Non-Markov local noise significantly slows decay, confirming  memory's coherence-preserving role. However, for the Werner-W state (i.e. $\rho_{W\!R}$), initial coherence increases with mixing parameter $p$. In common environments (Markov/non-Markov), coherence is preserved due to W-state stability. In Markov local environments, coherence decays exponentially, faster for higher $p$, with negligible coherence for $p=0.1$ while non-Markov local environments significantly slow down coherence decay.

\vskip 0.5cm
{\bf Declaration of competing interest} The authors declare that they have no known competing financial interests or
personal relationships that could have appeared to influence the work reported in this paper.

\vskip 0.5cm
{\bf Data availability statement} All data that support the findings of this study are included within the article. No supplementary file has been added.

\end{document}